\documentclass[vecphys]{svmult}
\usepackage{makeidx}         
\usepackage{graphicx}        
\usepackage{multicol}        
\usepackage{url}             
\usepackage[ps2pdf]{hyperref}  
\usepackage{float,amsmath}     
\usepackage{epsfig,epsf}
\usepackage{color}
\usepackage{graphicx}
\usepackage{epstopdf}
\newcommand{\beq}{\begin{equation}}
\newcommand{\eeq}{\end{equation}}
\def\eq#1{{(\ref{#1})}}

\makeindex

\begin{document}

{\Large\bf{In Search of the QCD--Gravity Correspondence}}
 
\vskip1cm

Theodor Bra\c{s}oveanu$^{1}$,  Dmitri Kharzeev$^{2}$ and Mauricio Martinez$^{3}$

\vskip1cm

1. Department of Physics, Princeton University \, \, \texttt{tbrasove@princeton.edu}

2. Department of Physics, Brookhaven National Laboratory \,\, \texttt{kharzeev@bnl.gov}

3. Helmholtz Research School and Otto Stern School, Goethe-Universit\"at Frankfurt am Main \, \,\texttt{guerrero@fias.uni-frankfurt.de}

\vspace{10 pt}
\begin{abstract}

Quantum Chromodynamics (QCD) is the fundamental theory of strong interactions. It describes the behavior of quarks and gluons which are the smallest known constituents of nuclear matter. The difficulties in solving the theory at low energies in the strongly interacting, non-perturbative regime have left unanswered many important questions in QCD, such as the nature of confinement or the mechanism of hadronization. In these lectures oriented towards the students we introduce two classes of dualities that attempt to reproduce many of the features of QCD, while making the treatment at strong coupling more tractable: (1) the AdS/CFT correspondence between a specific class of string theories and a conformal field theory and (2) an effective low-energy theory of QCD dual to classical QCD on a curved conformal gravitational background. The hope is that by applying these dualities to the evaluation of various properties of the strongly-interacting matter produced in heavy ion collisions one can understand how QCD behaves at strong coupling. We give an outline of the applications, with emphasis on two transport coefficients of QCD matter -- shear and bulk viscosities.

\end{abstract}

\section{Introduction.}

Recent results from the Relativistic Heavy Ion Collider (RHIC) at BNL reveal surprising dynamical 
properties of the strongly-coupled Quark-Gluon Plasma (sQGP). Some of these properties may be 
 explained through two less than traditional methods that employ dualities to re-formulate the underlying gauge theory in curved conformal gravitational backgrounds.  

\vspace{10pt}
The first approach is inspired by string theory and is based on the Anti-de-Sitter/ Conformal-Field-Theory (AdS/CFT) gravity/gauge theory duality~\cite{Mal9711, GKP9802, Wit9803}. This correspondence maps conformal strongly-coupled SU($N_{c}$) gauge theories (i.e. with large 't Hooft coupling $\lambda = g^2 N_c$) onto a weakly-coupled dual gravity theory. This formalism has been applied to conformal  gauge theories that share some features with QCD, such as a maximally super-symmetric $\cal N$=4 Yang-Mills theory. This leads to interesting predictions for several classes of observables that may be related to the QGP, such as entropy production, transport properties, jet quenching, dijet-bulk correlations etc (for a review, see \cite{Shuryak:2008eq}).

\vspace{10pt}
The second approach introduces an effective low-energy Lagrangian of QCD which satisfies the constraints imposed by the Renormalization Group, is scale and conformally invariant in the limit of vanishing vacuum energy density and matches the perturbative behavior at short distances (high energies)~\cite{KLT0403,Kharzeev:2008br}. This theory has a dual description as classical gluodynamics on a curved conformal space-time background on one hand, and gluodynamics in flat space-time coupled to scalar glueballs (which play the role of dilatons) on the other. In this approach, one may be able to describe confinement as an event horizon for colored particles \cite{Kharzeev:2005iz,CKS07}, in close analogy to what happens in the vicinity of a black hole. Recent efforts link the bulk viscosity of QCD matter and the associated entropy production to the breaking of scale invariance ~\cite{KT0705, KKT0711}.

\section{QCD and its properties.}

Quantum Chromodynamics (QCD) is at present universally accepted as the theory of the strong interaction. The fundamental degrees of freedom in the theory are the quarks and gluons that carry color charges. These particles cannot be directly seen in Nature, because confinement binds them into the color-neutral bound states -- baryons and mesons. QCD has been thoroughly tested in experiment and is known to possess the expected properties: the coupling constant weakens as the resolution scale grows (celebrated "asymptotic freedom" \cite{Gross:1973id,Politzer:1973fx}), the quarks and gluons manifest themselves through the production of jets, the partonic constituents of matter are seen in Deep Inelastic Scattering and the corresponding structure functions exhibit scaling violation, etc. 
\vskip0.3cm
However in spite of these successes the behavior of the theory at low energies or large distances and the structure of its vacuum state are still poorly understood. One may hope that a progress can be achieved through the studies of thermodynamics of quarks and gluons -- if history is any indication, 
understanding thermodynamic behavior may appear simpler than understanding the dynamics of the individual constituents. At temperatures accessible experimentally, the typical distances between the quarks and gluons in the "Quark-Gluon Plasma" (QGP) are quite large and the coupling is strong. Achieving a progress in the understanding of QGP thus requires methods that apply in the strong coupling domain.      
 In these lectures we will focus on two such methods, with applications to the transport coefficients of the quark-gluon plasma.

\subsection{The QCD lagrangian.}
QCD is formulated as a gauge theory, in analogy with QED. The structure of QED is entirely fixed by the requirement of invariance under local gauge transformations, that is invariance with respect to the phase rotation of the electron field $\exp (i\alpha (x))$, where the phase factor $\alpha$ depends on the space-time coordinates. In the case of QCD, we have the constituents -- quarks -- that come in three different colors, $N_c =3$. The local gauge invariance with respect to the $SU_c(N_c)$ rotations in color space introduces $N_c^2-1 = 8$ gauge bosons, the gluons.  Quarks are spin $1/2$ particles that belong to the fundamental representation of the $SU_c(3)$ whereas gluons are spin $1$ particles defined in the adjoint representation of $SU_c(3)$.  Since quarks can have three different colors -- say, red, green and blue -- a quark state vector can be expressed as a color multiplet of three components. Interactions with gluons re-paint the colors of the quarks; since color rotations do not commute (in other words, $SU_c(3)$ is a non-Abelian group) gluons can also interact with each other. 

\vspace{10pt}
Let us now formulate this mathematically. First,  consider the free lagrangian of quark fields:

\begin{equation}
\label{freelagr}
{\cal L}_{{\rm free}}=\sum_{q=u,d,s\dots}\sum_{colors}
\bar q(x)\left(\imag\gamma_\mu\frac{\partial}{\partial x_\mu}-m_q\right)q(x) 
\quad,
\end{equation}

and impose its invariance under a gauge transformation of the quark fields defined as

\begin{equation}
\label{gaugetrans}
U:\qquad q^j(x)\quad\to\quad U_{jk}(x)q^k(x).
\end{equation}

with j, k $\epsilon$ {1 . . .3} (we always sum over repeated indices). The fact that the lagrangian of the theory is invariant with respect to these gauge transformations implies that all physical observables must be gauge invariant.

\vspace{10pt} 
Note that in Eq. (\ref{gaugetrans}) $U$ is a unitary complex valued matrix, i.e., $UU^\dagger=U^\dagger U$=1 and $\det U$=1. In the fundamental representation of the group these matrices form the group $SU(3)$ with 3 being the number of colors $N_c$. This group has eight generators $T^a_{ij}$, $a \epsilon$ {1,2,..,8} of the fundamental representation of $SU(3)$, hence the matrix $U$ can be represented as

\begin{equation}
U(x)=\exp (-i\,\,\phi_a(x)\,T^a).
\end{equation}

The properties of $U$ imply that the generators $T^a_{ij}$ are Hermitian ($T^a=T^{a\dagger}$) and traceless ($Tr T^a$=0) (check this by making an infinitesimal transformation about unity). These generators satisfy a Lie algebra:

\begin{equation}
[T^a,T^b]=i f_{abc}T^c
\end{equation}

where $f_{abc}$ are $SU(3)$ structure constants. This means that unlike QED, QCD is a non-Abelian gauge theory: different color rotations do not commute.

\vspace{10pt} 

After we apply the local gauge transformation to the quark fields, the free lagrangian given in Eq. (\ref{freelagr}) acquires some extra terms proportional to $\partial_\mu\phi_a(x)$. In order to keep gauge invariance, it is necessary to compensate for this extra terms. This can be achieved by introducing the gauge field (in QCD, gauge fields will be understood as gluons) $A^\mu_{kj}$, and replacing the normal derivative $\partial_\mu$ in the free lagrangian (\ref{freelagr}) by the so-called covariant derivative:

\begin{equation}
\partial^\mu q^j(x)\quad\to\quad D^\mu_{kj}q^j(x)\equiv\left\{\delta_{kj}\partial^\mu
-\imag A_{kj}^\mu(x)\right\}q^j(x)
\end{equation}

Note that if we request for a gauge invariance, the lagrangian written in terms of covariant derivatives is no longer free, so we now have a coupling between quark fields and gauge fields. 
Under gauge transformations the gauge fields should transform as:

\begin{equation}
A^\mu(x)\to U(x)A^\mu(x)U^\dagger(x)
+\imag U(x)\partial^\mu U^\dagger(x).
\end{equation}

So the QCD lagrangian reads as\footnote{Hereafter we will omit color indices explicitly.}: 

\begin{equation}
\label{qcdlagr}
{\cal L}_{{\rm QCD}}=\sum_{q}\bar{q}(x)\left(\,i\,\gamma_\mu D^\mu-m_q\right)q(x)-\frac{1}{4g^2}{\rm tr}\;G^{\mu\nu}(x)G_{\mu\nu}(x)\quad
\end{equation}

where 

\begin{equation}
D_\mu=\partial_\mu -i\,A^a_\mu\,t^a;
\end{equation}
note that we have included $g$ in the definition of the gauge potential to reveal the dependence of the 
Lagragian \eq{qcdlagr} on the coupling constant; we will need it in what follows.

\vskip0.3cm

The first term of  (\ref{qcdlagr}) describes the dynamics of the interaction between the quarks and gluons while the second one describes the dynamics of the gluon field. 
The gluon field strength tensor is given by:

\begin{equation}
G^{\mu\nu}(x)\equiv{\rm i}\left[D^\mu,D^\nu\right]
=\partial^\mu A^\nu(x)-\partial^\nu A^\mu(x)
- i\,\left[A^{\mu}(x),A^{\nu}(x)\right]
\end{equation}

or in terms of the colour components $A^\mu_a$ of the gauge field:

\begin{equation}
\label{qcdstress}
G_a^{\mu\nu}(x)
=\partial^\mu A_a^\nu(x)-\partial^\nu A_a^\mu(x)
+f_{abc}A_b^\mu(x) A_c^\nu(x).
\end{equation}

The term $Tr(G^{\mu\nu}(x)G_{\mu\nu}(x))$ is also gauge invariant because 
$Tr\,(UG_{\mu\nu}G^{\mu\nu}U^{\dagger})\,=\,Tr\,(G_{\mu\nu}G^{\mu\nu})$. Note that the term $Tr\,(G_{\mu\nu}G^{\mu\nu})$ for the given stress tensor has non-linear couplings between the gauge fields themselves. Such self-interactions are responsible for the complexity of QCD dynamics. 

\subsection{Asymptotic freedom.}

One of the most remarkable properties of QCD is related to the fact that at large energies the coupling constant is small, i.e. perturbation theory is applicable. To understand better this aspect, let us first refer to what happens in QED. The electron-positron pairs screen the electric charge. Therefore, the electric charge becomes stronger at short distances. The dependence of the observed effective charge on the distance is given by:

\begin{equation}\label{run_qed}
e^2(r)=\frac{e^2(r_0)}{1+\frac{2e^2(r_0)}{3\pi}\log\frac{r}{r_0}}.
\end{equation}

This result can be obtained by resumming (logarithmically divergent, and regularized at the distance $r_0$) electron-positron loops dressing the virtual photon propagator. 
\vskip0.3cm
The running of the coupling constant with the distance prescribed by \eq{run_qed} has some paradoxical consequences. Indeed, at distances $r \gg r_0$ the coupling constant is seen to be 
independent of the value of the coupling at renormalization scale $r_0$. Moreover,  it vanishes as $\sim \log^{-1}(r/r_0)$ so
if we require the coupling constant be finite at $r_0$, in the local limit of $r_0 \to 0$ the coupling will vanish. This is 
so-called "Moscow zero" discovered by Landau and Pomeranchuk \cite{Landau:1955ip}. The possible ways out include the following: a) we know that QED is not a complete theory; at the scale $r_0 \sim 1/M_Z$ it has to be extended to the electroweak theory; b) At short distances QED is no longer weakly coupled, and so the perturbative expression \eq{run_qed} cannot be trusted and one should find a true non-perturbative answer.

\vspace{10pt} 
Instead of the dependence on the distance, we can also use momentum space and consider the dependence of the coupling on the virtuality of the photon, say, $q^2\equiv Q^2$. In terms of this variable, the expression \eq{run_qed} can be rewritten as:

\begin{equation}
\alpha (Q^2)=\frac{\alpha (Q_0^2)}{1-\frac{\alpha (Q_0^2)}{3\pi}\log\left(\frac{Q^2}{Q_0^2}\right)}
\end{equation}

The "Moscow zero" in momentum space manifests itself through the singularity at $Q^2 = Q_0^2 \exp(3 \pi / \alpha)$; note that since $\alpha \ll 1$, for all particle accelerator energies this pole is very far away and so QED is an excellent effective theory.  

\vskip0.3cm

In QCD the situation is different. Because of the gluon self-interactions, we have \textit{anti-}screening \cite{Gross:1973id,Politzer:1973fx}-- the constant coupling becomes small at short distances (high energies) but large at large distances (low energies).  In Coulomb gauge, the anti--screening stems from the diagram in which the exchange of a Coulomb gluon excites from the vacuum zero modes of the transverse gluons \cite{Khriplovich:1969aa,Gribov:1977wm,Zwanziger:2002sh}; for review see \cite{Dokshitzer:2004ie}. This diagram is purely real and leads to a Coulomb interaction that strengthens as the distance grows. As a result, the sign of $\log (Q^2/Q_0^2)$ changes. The final result is given by the expression:

\begin{equation}
\label{alphas}
\alpha_s(Q^2)=\frac{\alpha_s(Q_0^2)}{1+\frac{\alpha_s(Q_0^2)}{12\pi}(11\,N_C-2\,N_f)\,\log (Q^2/Q_0^2)}.
\end{equation}
\label{asymfree}

\begin{figure*}[t]
\centerline{
\includegraphics[width=6cm]{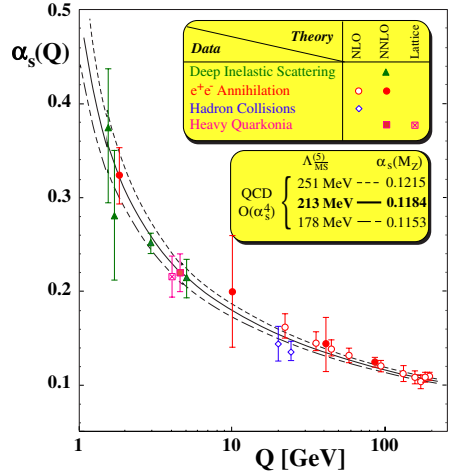}
}
\caption{
The running coupling constant $\alpha_s(Q^2)$ as a function
of momentum transfer $Q^2$ determined from a variety of processes; from \cite{Bethke}.
 	}
\label{fig:bethke}
\end{figure*}

In Fig. \ref{fig:bethke} we show the experimental verification of this prediction. Formally speaking, the fact that the coupling constant is small at high energies is related to the negative value of the so-called $\beta$-function. We will explain this in detail in Sect. \ref{breakscale}. 

\vskip0.3cm

Note that with \eq{alphas}, no singularity appears in the local limit -- so QCD by itself is a fully consistent field theory. However the pole is still present at small virtuality; it thus could affect all soft processes. This problem is very likely related to confinement of quarks and gluons, and finding the right way of dealing with it is akin to discovering the Holy Grail for the QCD theorists.

\subsection {Scale invariance breaking in QCD}
\label{breakscale}

Scale invariance plays an important role in many sub-fields of physics.  In a scale-inavriant theory, the physics looks the same at all scales. Suppose that   
 the action that describes the dynamics of some theory is invariant under dilatations:

\begin{equation}
x \to \lambda x
\end{equation}

If this is the case,  by Noether theorem, we have a dilatation current $s^\mu$ that is conserved and is given by:

\begin{equation}
s^\mu=x_\nu\,\theta^{\mu\nu}
\end{equation}

where $\theta^{\mu\nu}$ is the energy-momentum tensor. The conservation law reads as:

\begin{equation}
\partial_\mu s^\mu = \theta^\mu_\mu
\end{equation}

Therefore the divergence of the scale current corresponding to a scale transformation is equal to the trace of the energy-momentum tensor, and a scale-invariant theory will have $\theta_{\mu}^{\mu}$ = 0. Deviations from this will indicate a breaking of scale invariance.

\vspace{10pt} 
As a simple example consider classical electrodynamics without external sources: 

\begin{equation}
{\cal L}=-\frac{1}{4}F_{\mu\nu}F^{\mu\nu}
\end{equation}

>From this lagrangian, it is straightforward to obtain the energy-momentum tensor: 

\begin{equation}\label{four_dim}
\theta^\mu_\nu =-F^{\mu\rho}F_{\nu\rho}+\frac{1}{4}\delta^\mu_\nu F_{\rho\sigma}F^{\rho\sigma}
\end{equation}

Taking the trace of this tensor, we find that it will vanish  in four dimensions since in that case $\delta_\mu^\mu =4$; this shows that  classical electromagnetism without sources is scale invariant. Indeed, the result is the same at classical level for non-abelian Yang-Mills theories. 

\vskip0.3cm

However at quantum level this is not true. A simple way to understand this is the following: 
in quantum theory, the fluctuations can exist at all scales and thus their total energy is infinite. 
To obtain a finite result, we have to renormalize the theory. We do this at the cost of introducing the renormalization scale which 
is a dimensionful parameter. Clearly, its presence violates the original classical scale invariance of our theory.

\vspace{10pt}  
In Sect. \ref{asymfree}, we mentioned that the coupling constant in gauge theories changes as a function of the virtuality $Q^2$ due to quantum effects -- the fluctuations of the vacuum that dress the propagator of a gauge boson. Because of this, QED and QCD lose the invariance under scaling transformations at quantum level. Indeed, once quantum correction are taken properly, the trace of the energy-momentum tensor calculated from the QCD lagrangian (\ref{qcdlagr}) is~\cite{scale1,scale2}:

\begin{equation}
\label{dilcurr}
\partial^\mu s_\mu=\Theta^\mu_{\;\mu}=\sum_qm_q\bar qq+\frac{\beta(g)}{2g^3}
{\rm tr G^{\mu\nu}G_{\mu\nu}},
\end{equation}

where $\beta(g)$ is the QCD $\beta$-function, which governs the behavior of the 
running coupling: 

\begin{equation}
\label{rg}
\mu {d g(\mu) \over d \mu} = \beta (g). 
\end{equation}

As we already discussed in Sect. \ref{asymfree}, the small value for the coupling $\alpha_s$ at high energies is related to the sign of the $\beta$ function which is negative for QCD. This means that 
the theory is asymptotically free. The leading term in the perturbative expansion is (cf. Eq. (\ref{alphas}))

\begin{equation}
\label{beta}
\beta(g) = -b {g^3 \over (4\pi)^2} + O(g^5), \hskip1cm b = 11 N_c - 2 N_f, 
\end{equation}  

where $N_c$ and $N_f$ are the numbers of colors and flavors, respectively. 

\vspace{10pt} 
The hadron masses are defined as 
the forward matrix elements of trace of the QCD energy-momentum tensor: 
$2m_h^2=\left\langle  h|\Theta^\mu_{\;\mu}|h\right\rangle$ (the factor of $2m_h$ is due to relativistic normalization of states $|h\rangle$. The sum in \eq{dilcurr} runs over all quark flavors, and it might seem that heavy quarks can give a substantial contribution to the masses of light hadrons. However that does not happen since at small virtualities  the heavy flavor contribution to the sum (\ref{dilcurr}) is exactly canceled by a corresponding heavy flavor contribution to the $\beta$-function \cite{Shifman:1978bx}. 
Since the light quarks are light, the dominant contribution to the masses of most light hadrons comes from the gluon term in \eq{dilcurr} -- in other words, most of the observable mass in the Universe is due to the energy of gluon interactions.

\subsection{Confinement and the broken scale invariance.}

The asymptotic freedom allows us to probe quarks and gluons at short distances 
 when the coupling constant is small. However the growth of the coupling at large distances leads 
 to the binding of quarks and gluons into color-singlet hadrons of size $\sim 1$ fm.

\vspace{10pt} 
In quantum field theory, the evaluation of scattering amplitudes involves the concept of asymptotic states. However, the fact that quarks and gluons are confined prevents us from using quarks and gluons as asymptotic states within the $S$ matrix approach. An appropriate object for studying confinement of quarks is the so-called \textit{Wilson loop}~\cite{Wilson}:

\begin{equation}
\label{wiloop}
W(R,T) = Tr \left\{P\ exp\left[i\int_C A_{\mu}^a T^a dx^{\mu}\right]\right\},
\end{equation}

where $A_{\mu}^a$ is the gluon field and $T^a$ is the generator of $SU(3)$. Let us first choose  the contour $C$ in the integral as a rectangle with side $R$ in one of the space dimensions and $T$ in the time direction. With this contour we are dealing with the propagation of a heavy static quark and antiquark separated by a distance $R$ propagating in time for a period $T$. One important property of the Wilson loop is its relation to the potential acting between the static quark and antiquark: when $T\to\infty$,

\begin{equation}
\lim_{T \to \infty}\ W(R,T) = exp\left[- T V(R)\right],
\end{equation}

where $V(R)$ is the static potential between the heavy quarks. 

\vskip0.3cm

If we now decide to stretch the size of the rectangle in all directions by the same factor ($T\to\lambda T$ and $R\to\lambda R$ simultaneously) one can expect that the value of the Wilson loop will be modified. The only exception 
 is the Coulomb potential, which maintains the scale invariance in the asymptotic value of the Wilson loop:

\begin{equation}
W(R,T)=W(\lambda\,R,\,\lambda\,T), \hspace{0.2cm} \texttt{if} \hspace{0.2cm} V(R)\sim 1/R.
\end{equation}

For QCD, the quark-antiquark potential has the form

\begin{equation}
V(R) = -\frac{4}{3}\frac{\alpha_s (R)}{R}+\sigma R, \label{string}
\end{equation}

where $\sigma$ is the tension of the string stretched between the quark and antiquark. Phenomenology tells us that its value is around $\sim 1 \ {\rm GeV}/{\rm fm}$. As one can easily see, both the running coupling and confinement are in violent contradiction with the  scale invariance.

\vspace{10pt} 
In the limit when $T\to\infty$ and large distances, the asymptotic behavior of the Wilson loop reads approximately as:

\begin{equation}
W(R,T) \simeq exp\left[- \sigma T R\right], \label{area}
\end{equation}

This is the famous "area law" of the Wilson loop that signals confinement. 

\vskip0.3cm

At finite temperature, various transport properties of QCD matter appear sensitive to confinement.   
 In Sect. \ref{bulkvisc} we discuss the intriguing relation between bulk viscosity and breaking of scale invariance.

\section {Black holes.}

The concept of a black hole dates back to the 18th century, when the British geologist and astronomer Rev. John Mitchell advanced the idea of the existence of a body so massive that the escape velocity at its surface would be equal to the speed of light. In a paper he wrote to the Royal Society in 1783 he concludes that "\textit{all light emitted would be made to return towards it, by its own proper gravity.}" This was due in part to the popularity of the "corpuscular theory" of light at the time, which made possible that light could be affected by gravity in the same way as ordinary matter~\cite{Ell99}.

\vspace{10pt}
The same idea of strong-gravity objects trapping light was proposed by the mathematician Pierre-Simon Laplace in his book \textit{Exposition du Systeme du Monde} in 1796, though it did not gain much consideration throughout the 19th century, as the "wave theory" of light was gaining ground.  

\vspace{10pt}
Later on, upon the advent of general relativity as a new theory of gravity in 1915 and the formulation of Einstein's field equations, relating the curvature of spacetime with the matter and energy content within the spacetime (via the stress-energy tensor $T_{\mu\nu}$)

\begin{equation}
G_{\mu\nu} = \frac{8\pi G}{c^4}T_{\mu\nu},    
\end{equation}

the German physicist and astronomer Karl Schwarzschild gave the solution for a static, isotropic gravitational field, predicting the existence of a singularity at r=2GM, the so-called Schwarzschild radius~\cite{Sch16}. Schwarzschild introduced a metric

\begin{equation}
ds^2 = c^2 dt^2 (1 - \frac{2GM}{rc^2}) - \frac{dr^2}{1 - \frac{2GM}{rc^2}} - r^2 d\Omega^2
\end{equation}

which was proved to be the most general solution of Einstein's equations without matter ($T_{\mu\nu} = 0$) and spherical symmetry -- via Birkhoof's theorem in 1923. It was later used by Oppenheimer and Volkoff in 1939 to predict the collapse of massive  stars~\cite{Opp39}. It was not until the introduction of the Eddington-Finkelstein coordinates in 1958 that the Schwarzschild surface r=2GM was interpreted as the event horizon of a black hole, acting as a perfect unidirectional membrane. That is, information from outside the boundary was allowed to cross it, but events occurring inside the boundary could in no way affect an outside observer~\cite{Fin58}.

\vspace{10pt}
A major breakthrough came in the 1970s, when Jacob Bekenstein proposed that a black hole should have a finite non-zero entropy and temperature, proportional to the area of its horizon. A finite entropy is in agreement with the second law of thermodynamics. By applying quantum field theory to the curved spacetime around the event horizon, Stephen Hawking concluded in 1974 that black holes should emit thermal radiation~\cite{Haw74}. He found the temperature of the emitted radiation to be proportional to the acceleration due to gravity of a near-horizon observer

\begin{equation}
\label{BHtemp}
T_{BH} = \frac{\hbar c^3}{8\pi GMk_{B}}
\end{equation}

and confirmed Beckenstein's conjecture by fixing the constant of proportionality between the entropy and the area of the black hole event horizon. In the equation above, $k_B$ is Boltzmann's constant, G is Newton's gravitational constant, h is Plank's constant, while M is the mass of the black hole. 

\vskip0.3cm

Interestingly, in gravitational theories the black hole entropy appears to be the maximum entropy that one can squeeze within a fixed volume, and this entropy is proportional to the area $A$ of the black hole horizon:

\begin{equation}
\label{BHentropy}
S_{BH} = \frac{k_{B}Ac^3}{4G\hbar},
\end{equation}

This is in sharp contradiction to the "usual" physics when the entropy is proportional to the volume of the system. Since the entropy counts the number of degrees of freedom, one may wonder whether 
the "true" number of degrees of freedom in gravitational theory is smaller than naively expected, and whether the dynamics might be formulated as a dynamics of the surface modes. We will come back to this crucial question shortly.

 \vskip0.3cm
 
 An interesting consequence of~(\ref{BHtemp}) is that black holes that are less massive than the planet Mercury would become hotter than the cosmic microwave background (about 2.73 K) and would slowly evaporate with time by giving up energy through Hawking radiation. As their mass decreased, their temperature would gradually increase. Thus, small black holes will eventually undergo runaway evaporation and vanish in a burst of radiation.

\vspace{10pt}
During the last decade, concerns regarding the formation in high-energy particle accelerators of black holes that would accrete ordinary matter and put in danger the Earth were firmly dismissed. 
The reader can easily check on the basis of formulae given in this section that the energies of 
any existing or planned accelerator is many orders of magnitude below the one needed for 
the black hole formation
 in either classical or quantum gravitational framework. The corresponding studies have been done at both the Relativistic Heavy-Ion Collider (RHIC)~\cite{Bus99} and more recently at the Large Hadron Collider (LHC). In the case of RHIC, the speculative disaster scenarios were to some extent due to the misinterpretation of a mathematically dual treatment of the hot QCD matter as a black hole in $AdS_{5}$ x $S^{5}$ space via the AdS/CFT correspondence~\cite{Nas05}, which is the topic of section \ref{AdSCFT}.

\section {Using black holes to understand gauge theories.}

\subsection {The AdS/CFT correspondence.}
\label {AdSCFT}

The AdS/CFT correspondence is a successfully tested conjecture about the mathematical equivalence of a string theory defined on an Anti de Sitter (AdS) space and a conformal field theory\footnote{Conformal invariance is a generalization of scale invariance.} defined on the boundary of the AdS space. It follows as an application of the holographic principle~\cite{Bous02}, developed by t'Hooft and Susskind, which states that the description of a volume of space can be encoded on a boundary to that region of space\footnote{In case of a black hole, the holographic principle states that the description of all objects falling into the black hole is entirely given by surface fluctuations of the event horizon.}.

\vspace{10pt} Here we focus on a particular version of this equivalence, namely the duality between Type IIB string theory on $AdS_{5}$ x $S^{5}$ and a supersymmetric $\cal N$=4 Yang-Mills gauge theory\footnote{$\cal N$=4 supersymmetric Yang-Mills is a conformal field theory.} on the 4-dimensional boundary of $AdS_{5}$, as it was originally introduced by Maldacena~\cite{Mal9711} in 1997, followed by Gubser, Klebanov, Polyakov~\cite{GKP9802} and Witten~\cite{Wit9803} in 1998.

\subsubsection{Anti de Sitter Space.}

On the gravity side, the requirement of conformal invariance fixes the metric of the $5^{th}$ dimension uniquely; it is an Anti de Sitter space $AdS_{5}$ -- that is, a space with Lorentzian signature (- + +...+) and constant negative curvature~\cite{Nas07}. The Anti de Sitter space is a maximally symmetric, vacuum solution of Einstein$'$s field equations with negative cosmological constant $\Lambda$ $<$ 0. In d dimensions, it is defined by an embedding in d+1 dimensions 

\begin{equation}
ds^2 = - dx_{0}^2 + \sum_{i=1}^{d-1}{dx_{i}^2} - dx_{d+1}^2,
\end{equation}

\begin{equation}
\label{R_AdS}
-R^2 = - x_{0}^2 + \sum_{i=1}^{d-1}{x_{i}^2} - x_{d+1}^2,
\end{equation}

which makes it the Lorentzian version of Lobachevski space. It is invariant under the group SO(2,d-1) that rotates the coordinates $x_\mu = (x_{0}, x_{d+1}, x_{1}, ..., x_{d-1})$ into $x'^{\mu} = \Lambda^{\mu}_{\nu} x^\nu$. In Poincar$\acute{e}$ coordinates, the metric of this space is given by

\begin{equation}
ds^2 = \frac{R^2}{x_{0}^2}(- dt^2 + \sum_{i=1}^{d-2}{dx_{i}^2} + dx_{0}^2),
\end{equation}

where $t>-\infty$, $x_{i}<+\infty$ and $0<x_{0}<+\infty$. Up to a conformal factor, this is just
like (flat) 3d Minkowski space, though not all space is covered. Thus, in these Poincar$\acute{e}$ coordinates, Anti de Sitter space can be understood as a d-1 dimensional Minkowski space in $(t, x_{1}, ..., x_{d-2})$ coordinates, with a \textit{warp factor} (gravitational potential) that depends only on the additional coordinate $x_{0}$.

\subsubsection{Supersymmetry.}

Now, our conformal field theory needs to be supersymmetric -- that is, associate a fermion (particle with half-integer spin) to every type of boson (particle with integer spin) and a boson to every type of fermion. This can be represented by a graded Lie algebra generalization of the Poincar$\acute{e}$ + internal symmetries, with ''even'' generators $P_a$, $J_{ab}$, $T_r$ and ''odd'' generators $Q^i_\alpha$, satisfying the following commuting and anticommuting laws~\cite{Nas07}:  

\begin{equation}
\left[even, even\right] = even; \,\left\{odd, odd\right\} = even; \,\left[even, odd\right] = odd.
\end{equation}

We note that $P_a$ and $J_{ab}$ are, respectively, the generators of 3+1 dimensional translation symmetries and the Lorentz generators of the SO(1,3) Lorentz group, which together define the Poincar$\acute{e}$ symmetry, $T_r$ correspond to internal symmetries of particle physics such as local U(1) of electromagnetism or local SU(3) of QCD or global SU(2) of isospin, while $Q^i_\alpha$ are spinors satisfying the supersymmetry algebra

\begin{equation}
Q\,boson = fermion; \,Q\,fermion = boson.
\end{equation}

Since all symmetries of the field theory should be reflected in the dual string theory, the supersymmetry requirement further constrains the 10-dimensional string theory to live in $AdS_{5}$ x $S^{5}$ (that is 5-dimensional AdS space times a 5-sphere)~\cite{AGMOO99}. We make sense of the higher dimensions ($D > 4$) of the theory by means of Kaluza-Klein dimensional reduction, which states that the extra D -- 4 dimensions are all curled up in a small space, in the form of a small sphere or torus.

\subsubsection{D Branes and Black Holes in String Theory.}

The metric commonly used in AdS/CFT calculations is the $AdS_{5}$-Schwarzschild solution, which describes the near-horizon geometry of D3-branes (see below): 

\begin{equation}
\label{AdS_Schwarz}
ds_{5}^2 = \frac{L^2}{z^2}\left[- h(z)dt^2 + d\stackrel{\rightarrow}{x}^2 + \frac{dz^2}{h(z)}\right], \\h(z) = 1 - \frac{z_{H}^4}{z^4}.
\end{equation}

\begin{figure}
\centering
\includegraphics[width=8cm]{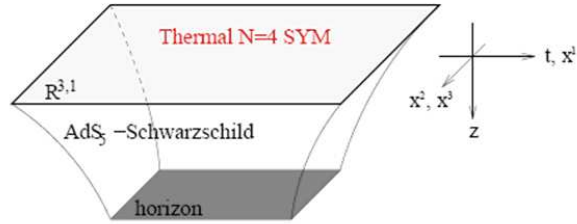}
\caption{The AdS-Schwarzschild solution, dual to a thermal state of $\cal N$=4 super-Yang-Mills, by Gubser.}
\label{fig:AdS_Schwarz}
\end{figure}

Here L is the common radius of curvature of $AdS_5$ and $S^{5}$, $z = \frac{L^2}{r}$ and $z_H$ is the corresponding z-coordinate of the black-hole horizon\footnote{z ranges from 0 to $\infty$ and corresponds to what we previously defined as $x_0$ in~(\ref{R_AdS})}. This metric extremizes an action derived from type IIB string theory on $S^{5}$:

\begin{equation}
S = \frac{1}{16\pi G_{5}}\int d^{5}x\sqrt{-g}(R+\frac{12}{L^2}).
\end{equation}

\vspace{10pt}
The relationship between gauge theories and string theory on Anti-de-Sitter spaces was motivated by studies of D-branes and black holes in string theory~\cite{AGMOO99}. D-branes are solitons in string theory and are defined (in string perturbation theory) as surfaces where open strings can end. These open strings have massless modes describing the oscillations of the branes, a gauge field living on the brane or their fermionic partners. Having N coincident branes on which the open strings can start and end points us towards a low energy dynamics described by a U(N) gauge theory. Now, if one considers D-p-branes, which are charged under p+1-form gauge potentials, and adds to them D-branes, they generate together a flux of the corresponding field strengths associated with the gauge potentials, which contributes to the stress energy tensor; so the geometry becomes curved. One can find solutions of the supergravity equations carrying these fluxes. These solutions are very similar to extremal charged black hole solutions in general relativity, except that in this case they are black branes with p extended spatial dimensions (which, like black holes, contain event horizons).

\begin{figure}
\centering
\includegraphics[width=8cm]{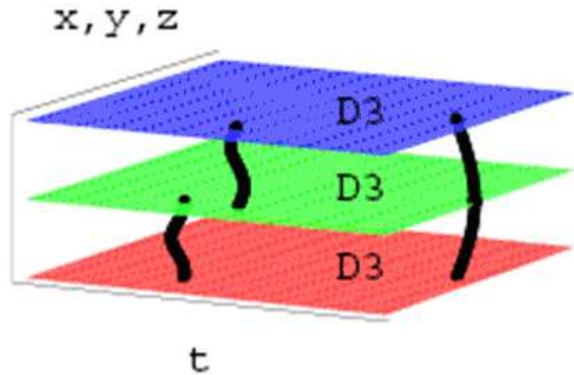}
\caption{D-brane representation, by Gubser.}
\label{fig:D_Branes}
\end{figure}

\vspace{10pt}
If one considers a set of N coincident D-3-branes, the near horizon geometry turns out to be the famed $AdS_{5}$ x $S^{5}$~\cite{AGMOO99}. We also know that the low energy dynamics on their worldvolume is governed by a U(N) gauge theory with $\cal N$=4 supersymmetry. These two pictures of D-branes are perturbatively valid for different regimes of the coupling. While perturbative field theory is valid when $g_{s}$N is small, the low-energy gravitational description is perturbatively valid when the radius of curvature is much larger than the string scale, that is when $g_{s}$N becomes very large. As an object is brought closer to the black brane horizon, its energy measured by an outside observer gets redshifted by the large gravitational potential, so it becomes very small. Since low energy excitations on the branes are governed by the Yang-Mills theory, it becomes natural to conjecture that Yang-Mills theory at strong coupling is describing the near horizon region of the black brane, whose geometry is $AdS_{5}$ x $S^{5}$! 

\vspace{10pt}
Near r=0, the extremal 3-brane geometry given in~(\ref{AdS_Schwarz}) is non-singular and
all appropriately measured curvature components become small for large L. Thus, for L much larger than the string scale $\sqrt{\alpha'}$, the entire 3-brane geometry has small curvature everywhere and can be described by the supergravity approximation to type IIB string theory. By expressing the ADM tension (mass per unit area) of the extremal 3-brane classical solution to N times the tension of a single D3-brane, one obtains the relation~\cite{Kle00}:

\begin{equation}
\label{ADM_tension}
\frac{2}{\kappa^2}L^4\Omega_5 = N\frac{\sqrt{\pi}}{\kappa}.
\end{equation}

In this context, $\Omega_5 = \pi^3$ is the volume of a unit 5-sphere and $\kappa = \sqrt{8\pi G}$ is the 10-dimensional gravitational constant. Since $\kappa = 8\pi^{7/2}g_s \alpha'^2$ and $g_{YM}^2 = 4\pi g_s$, Eq.~(\ref{ADM_tension}) becomes

\begin{equation}
L^4 = g_{YM}^2 N {\alpha'}^2,
\end{equation}

where $g_s$ and $g_{YM}$ are the string and Yang-Mills couplings, respectively. Thus, for large $L \gg \sqrt{\alpha'}$ a strong t'Hooft coupling $g_{YM}^2 N \gg 1$ is required. This remarkable result lies at the heart of the success of the AdS/CFT correspondence. We further point to references~\cite{Nas07, AGMOO99, Kle00} for comprehensive reviews of the conjectured equivalence.

\subsection {QCD and Gravity}
\label {QCDGrav}

As we discussed above, asymptotic freedom and confinement explicitly break conformal invariance in QCD, and the AdS/CFT correspondence relies on conformal symmetry in a very essential way. This is 
at present the main obstacle in the way of applying these ideas to QCD. 
We are interested in the strong coupling behavior which in QCD is encountered in the 
 the low energy regime. It is thus natural to try and gain a physical insight on confinement by considering effective low-energy theories. In this section we will sketch the construction of one of these theories. What is interesting in this example is that it also points toward gravity: the resulting effective lagrangian can be mathematically reformulated as classical QCD on a curved conformal gravitational background. Our discussion follows the arguments from Ref.~\cite{KLT0403}.

\vspace{10pt}
We would like to construct an effective theory that is not scale invariant but transforms under scale transformations in a well-defined way prescribed by the asymptotic freedom, i.e. is invariant under the  \textit{Renormalization Group (RG)} transformations\footnote{Renormalization group is a field theory technique that allows one to investigate the changes of a physical system as one views it at different distance scales.}. 
Since as we have seen above the scale transformations are generated by the trace of the energy momentum tensor, we will begin by encoding the dynamics prescribed by the RG 
into a set of low energy theorems for the correlation functions of the trace of the energy-momentum tensor. We reproduce the derivation \cite{NSVZ81} of these theorems here because they will be relevant in the construction of the effective lagrangian.

\vspace{10pt}
The expectation value of an operator ${\cal O}$ that is invariant with respect to the change of renormalization scale $M_0$ can be written as

\begin{equation}
\label{operator}
\langle {\cal O}\rangle\,\sim\, \left[ M_0\,e^{-\,\frac{8\,\pi}
{b\, g^2(M_0)}}\right]^d;
\end{equation}
this way the expectation value does not change with the scale $M_0$, and there  is thus no \textit{anomalous dimension}, only the canonical dimension $d$.

On the other hand, we can write down this expectation value as a functional integral that will contain the exponential of the action. We have seen above (see \eq{qcdlagr}) that the action density in gluodynamics depends on the coupling constant $g$; it is proportional to $1/4g^2$:  

\begin{equation}
{\cal L}\,=\,(-1/4g^2)\tilde F^{a\mu\nu}\tilde F^a_{\mu\nu}
\end{equation}

where $\tilde F = g F$ is the rescaled gluon field. The derivative w.r.t. $-1/4g^2(Q^2)$ of the expectation value of ${\cal O}$ thus generates a zero-momentum correlation function:

\begin{equation}
\label{tracerelation}
i\,\int\,dx\,\langle T\{\,{\cal O}(x)\,,\, \tilde F^2(0)\,\}\rangle\,=\,
-\frac{d}{d(-1/4g^2)}\,\langle {\cal O}\rangle;
\end{equation}
this procedure is analogous to differentiating w.r.t. the inverse temperature in statistical mechanics. 
Combining Eqs.~(\ref{operator}) and~(\ref{tracerelation}), we obtain \cite{NSVZ81} for $d=4$:

\begin{equation} 
\label{twopoint}
i\,\lim_{q\rightarrow 0}\,\int dx\,e^{i\,q\,x}\,\langle0|T\{\,{\cal O}(x)\,,\,
\frac{\beta(\alpha_s)}{4\,\alpha_s}\, F^2(0)\,
\}|0\rangle_\mathrm{connected}=
\langle O\rangle\, (-4)\,.
\end{equation}

This result can be easily generalized through an iteration method to obtain a set of relations between n-point correlation functions and an arbitrary number of operators $F^2$. If our operator ${\cal O}$ is the trace of the energy-momentum tensor itself, i.e.

\begin{equation}
{\cal O}\equiv \theta^\mu_\mu=\frac{\beta (g)}{2\,g} F^a_{\mu\nu}\, F^{a\mu\nu},
\end{equation}

then Eq.~(\ref{twopoint}) can be written as:

\begin{align}
\label{yangscale}
i^n
\int dx_1\ldots dx_n\,\langle0|T\{\theta_{\mu_1}^{\mu_1}(x_1),
\ldots, & \theta_{\mu_n}^{\mu_n}(x_n),\theta_{\mu}^{\mu}(0)
\}|0\rangle_\mathrm{connected}\\ \nonumber
&=\,
\langle \theta_{\mu}^{\mu}(x)\rangle\, (-4)^n.
\end{align}

This infinite chain of low-energy theorems determines the structure of the low-energy theory completely, and we will now
construct our effective lagrangian using the method developed in \cite{Migdal:1982jp}.

\vskip0.3cm

Let us consider gluodynamics in a curved conformally flat background in $d$ dimensions, with a background described by the metric:

\begin{equation}\label{metric}
g_{\mu\nu}(x)\,=\,e^{h(x)}\, \delta_{\mu\nu}.
\end{equation}

The action of the gluon fields in this curved background is:

\begin{equation}
\label{action}
S\,=\,-\frac{1}{4\, g^2}\,\int d^dx\,\sqrt{-\,g} \,
g^{\mu\nu}\, g^{\lambda\sigma}\, \tilde F^a_{\mu\lambda}\,
\tilde F^a_{\nu \sigma},
\end{equation}

with $g=\det\,g^{\mu\nu}$. Note that Yang-Mills theory on a curved background is scale and conformally invariant in any number of dimensions $d$, contrary to the case a flat space when the same theory is scale and conformally invariant only if $d$=4, see \eq{four_dim}. The regularization of the action brings in an extra term in Eq.~(\ref{action}) in $d=4$:

\begin{equation}
\label{anoma}
\Delta S\,=\,-\frac{1}{4\, g^2}\, \int d^4x\, e^{2h}\,
\left[-\frac{b\,g^2}{32\,\pi^2}(\tilde F^a_{\mu\nu})^2\right].
\end{equation}

where $b=11\,N_C/3$. This new term is proportional to the second term of the right hand side of Eq.~(\ref{dilcurr}). Therefore, the explicit breaking of the scale invariance of QCD manifests itself in the theory defined by the effective action given by the sum of Eqs. (\ref{action}) and (\ref{anoma}) through a term containing the auxiliary scalar field $h(x)$~\cite{Migdal:1982jp}, without any dimensionful parameters. For a theory defined on a flat space-time, the scale anomaly presents itself in the phenomenon of dimensional transmutation, which brings in a dimensionful parameter explicitly.

\vspace{10pt}
The kinetic part for the field $h(x)$ can be obtained in a manifestly scale and conformally
invariant way using the Einstein-Hilbert Lagrangian for the one-loop effective Yang-Mills
field:

\begin{equation}
\label{effact}
S_{eff}\,=\,\int d^4x\,\sqrt{-g}\left( \frac{1}{8\,\pi\, G}\, R
\,-\frac{1}{4\, g^2} \,
g^{\mu\nu}\, g^{\lambda\sigma}\, \tilde F^a_{\mu\lambda}\,
\tilde F^a_{\nu \sigma}\,-\, 
e^{2h}\,\theta_\mu^\mu
\right),
\end{equation}

where  $G$ is some dimensionful constant and $R$ is the Ricci scalar. Using \eq{metric} we get

\beq\label{ricci}
R \sqrt{-g} \equiv R_{\mu}^{\mu} \sqrt{-g} = e^h \frac{3}{2} (\partial_\mu h)^2.
\eeq
In deriving \eq{ricci} we have neglected terms of higher order in derivatives and constrained ourselves to the Einstein's gravity. This corresponds to an expansion in powers of the gradients for a slowly varying background field.

\vspace{10pt}
With this example, we have constructed an effective lagrangian that preserves scale invariance of QCD in a classical curved background at a price of a new dilaton field $h(x)$. 

\subsubsection{Confinement as an event horizon for colored particles?}

In the previous section we have constructed an effective low-energy lagrangian that preserves the scale invariance. This effective theory is mapped mathematically onto classical QCD on a curved conformal gravitational background. It is interesting that this theory may offer a geometrical way of associating confinement with an event horizon for colored particles \cite{CKS07}.
 In general relativity the appearance of an event horizon occurs through the modification of the underlying space-time structure by the gravitational interaction. Our effective lagrangian ${\cal L}$ depends on a single background field, and the modification of the metric can be computed as follows~\cite{Novello:1999pg}:

\begin{equation}\label{effmetric}
g_{\mu\nu}=\eta_{\mu\nu}{\cal L}'-4\,F_{\alpha\mu}F^\alpha_\nu{\cal L}''
\end{equation}

where the primes indicate first and second derivatives with respect to $F\equiv F_{\mu\nu}F^{\mu\nu}$. The vanishing of the temporal component of the modified metric will define the compact region of the theory, i.e. the counterpart of a black hole. It is possible to write the effective lagrangian derived from the action (\ref{effact}) as:

\begin{equation}
\label{efflagran}
{\cal L}_{QCD} = {1\over 4} F_{\mu \nu}F^{\mu \nu} {g^2(0) \over g^2(gF)}
=\frac{1}{4} F_{\mu \nu}F^{\mu \nu} {\epsilon(gF)},
\end{equation}

with the "dielectric" constant of the system under the presence of the background field at one loop given by:

\begin{equation}
{\epsilon(gF)} \simeq 1 - \beta_0 \left({g^2 \over 4\pi}\right)
\ln {\Lambda^2 \over gF}.
\end{equation}

Here $\beta_0=(11N_c - 2N_f)/48\pi^2$, while $N_c$ and $N_f$ specify the number of colors and flavors, respectively. 
Therefore the effective metric \eq{effmetric} computed for the 
lagrangian~(\ref{efflagran}) can yield the vanishing temporal component $g_{00}=0$ when

\begin{equation}
gF^* = \Lambda^2 \exp\{-4\pi/\beta_0 g^2\}.
\end{equation}

Therefore, an event horizon may be formed at $r^* \sim 1/\sqrt{gF^*}$ in our effective theory -- it would take colored particles an infinite time to escape. In black hole physics there is a deep relation between the Hawking temperature and the existence of an event horizon. Consequently, the radiation emitted by a black hole and its evaporation process might be related to thermal hadron production (see \cite{CKS07} for details).

\section {Applications: shear and bulk viscosities.}

To exemplify the use of the two methods presented previously in sections ~\ref{AdSCFT} and~\ref{QCDGrav} we will discuss the calculation of two important transport coefficients of QCD matter, which can be linked to relevant observables in studies of ultrarelativistic heavy-ion collisions.

\subsection {Definitions.}

In finite temperature field theory, the shear and bulk viscosity are defined as transport coefficients of the energy-momentum tensor. For this, we consider a plasma slightly out of equilibrium, such that there is local thermal equilibrium everywhere, but the temperature and average velocity are allowed to slowly vary in space. Then one can define at any point a local rest frame, where the 3-momentum density vanishes, $\theta_{i0} = 0$. In this frame, one has the following constitutive relation for the energy-momentum tensor~\cite{PSS01, ADM06}:

\begin{equation}
\theta_{ij} = P_{eq}(\epsilon)\delta_{ij} - \eta(\partial_i u_j + \partial_j u_i - \frac{2}{3}\delta_{ij}\partial_k u_k) - \zeta \delta_{ij}\stackrel{\rightarrow}{\nabla} \cdot \stackrel{\rightarrow}{u},
\end{equation}

where $P_{eq}(\epsilon)$ is the pressure (related to the energy density of the system through an equilibrium equation of state), $\eta$ is the shear viscosity, $\zeta$ the bulk viscosity and $u_i$ are the flow velocities. All kinetic coefficients can be expressed, via Kubo formulas, as static limits of correlation functions of the corresponding currents, namely~\cite{LP80}:

\begin{equation}
\label{eta_Kubo}
\eta = \frac{1}{2}\lim_{\omega\to 0}\frac{1}{\omega}\int_0^\infty dt \int d^3r\,e^{i\omega t}\,\langle [\theta_{xy}(t,\textbf{x}),\theta_{xy}(0,\textbf{0})]\rangle,
\end{equation}

\begin{equation}
\label{zeta_Kubo}
\zeta = \frac{1}{9}\lim_{\omega\to 0}\frac{1}{\omega}\int_0^\infty dt \int d^3r\,e^{i\omega t}\,\langle [\theta_{ii}(t,\textbf{x}),\theta_{kk}(0,\textbf{0})]\rangle.
\end{equation}

The shear viscosity $\eta$ indicates how much entropy is produced by transformation of shape at constant volume; it is generated by translations. Correspondingly, the bulk viscosity $\zeta$ quantifies how much entropy is produced by transformation of volume at constant shape; it is generated by dilatations.

\subsection {The shear viscosity bound.}
\label{shearvisc}

In~\cite{KSS04}, the ratio of shear viscosity to volume density of entropy $\eta/s$ is computed for a large class of strongly interacting quantum field theories which have a dual description involving black holes in Anti de Sitter space. This ratio can be used to characterize how close a given fluid is to being perfect.

\vspace{10pt} 
A first step is to relate the shear viscosity as defined in Eq.~(\ref{eta_Kubo}) to the absorption cross-section of low-energy gravitons. We consider a graviton of frequency $\omega$, polarized in the xy direction and propagating perpendicularly to a brane. Since according to the gauge-gravity duality, the stress-energy tensor couples to the metric perturbations at the boundary~\cite{AGMOO99}, then in the field theory picture the absorption cross-section of the graviton by the brane is proportional to the imaginary part of the retarded Green function of the operator coupled to $h_{xy}$, that is $\theta_{xy}$:

\begin{equation}
\label{sigma_abs}
\sigma_{abs}(\omega) = -\frac{2\kappa^2}{\omega} Im G^R(\omega) = \frac{\kappa^2}{\omega} \int_0^\infty dt \int d^3r\,e^{i\omega t}\,\langle [\theta_{xy}(t,\textbf{x}),\theta_{xy}(0,\textbf{0})]\rangle,
\end{equation}

where $\kappa = \sqrt{8\pi G}$ emerges as a result of the normalization of the graviton's action. Upon comparing Eq.~(\ref{eta_Kubo}) to~(\ref{sigma_abs}), one obtains immediately the shear viscosity as a function of the graviton absorption cross-section:

\begin{equation}
\eta = \frac{\sigma_{abs}(0)}{2\kappa^2} = \frac{\sigma_{abs}(0)}{16\pi G}.
\end{equation}

The next step is to prove that the absorption cross-section of a graviton by a black hole is the same as that of a scalar, which is equal in the low-frequency limit $\omega \rightarrow 0$ to the area of the horizon, namely $\sigma_{abs}(0) = a$.

\vspace{10pt}
Finally, one needs to compute the volume density of entropy. Again, we use the dual holographic description and conclude that the entropy of the dual field theory is equal to the entropy of the black brane, which is proportional to the area of its event horizon, as found by Beckenstein and Hawking in Eq.~(\ref{BHentropy}). The entropy density is therefore

\begin{equation}
s = \frac{k_{B}a}{4G\hbar},
\end{equation}

which results in a shear viscosity to entropy density ratio of

\begin{equation}
\label{eta_s}
\frac{\eta}{s} = \frac{\hbar}{4\pi k_{B}}.
\end{equation}

It is remarkable that the result -- which applies in the limit of strong coupling, where the gravity description is appropriate -- does not depend on the specific metric chosen, the ratio being the same for various types of Dp- and Mp-branes, even if the corresponding dual theories are very different. Furthermore, according to a conjecture due to the authors of~\cite{KSS04}, this result can be viewed as a universal lower bound for all relativistic quantum field theories at finite temperature and zero chemical potential. The inequality is to be saturated by theories with gravity duals. We further refer to~\cite{KSS04} for the arguments supporting this conjecture; see also \cite{Kats:2007mq,Buchel:2008vz} for a discussion of circumstances under which this bound may be violated.

\vspace{10pt}
Recent lattice results \cite{Mey0704} give $\eta/s=0.134(33)$ for $T=1.65\,T_c$ and $\eta/s=0.102(56)$ for $T =1.24\,T_c$ -- the values which are above, but not much above, the bound~(\ref{eta_s}).

\begin{figure}
\centering
\includegraphics[width=6cm]{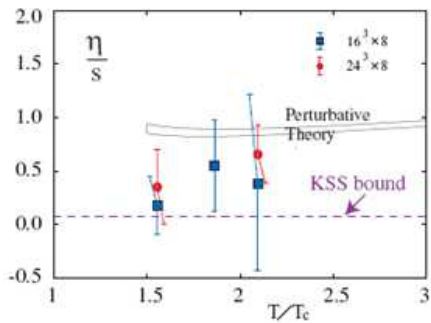}
\caption{The ratio of the shear viscosity to the entropy as a function of T/$T_c$ from \cite{NS04}. KSS bound is 1/4$\pi$.}
\label{fig:Eta_lat}
\end{figure}

\vspace{10pt}
At the same time, RHIC data provide a limit for the ratio $\eta/s$ by measuring the elliptic flow of particles produced in very energetic heavy-ion collisions. Results so far have pointed towards a very low-viscosity, nearly perfect fluid of quarks and gluons \cite{Dusling:2007gi,Romatschke:2007mq,Luzum:2008cw,Song:2008hj}. 

\subsection {Bulk viscosity and hadronization.}
\label{bulkvisc}

The transport coefficient of the plasma which is directly related to its conformal properties is the bulk viscosity~\cite{KT0705}. Indeed, it is related by Kubo's formula to the correlation function of the trace of the energy-momentum tensor. It is clear from Eq.~(\ref{zeta_Kubo}) that for any conformally invariant theory with $\theta^\mu_\mu=0$ the bulk viscosity should vanish. This is the case for $\cal N$=4 SUSY Yang-Mills theory. In contrast, a non-zero $\zeta$ can be generated by breaking the scale invariance.

\vspace{10pt}
A perturbative evaluation by~\cite{ADM06} has yielded very small values for $\zeta$, with the ratio $\zeta$/$\eta$ of the order of $10^{-3}$ for 3-flavor QCD with $\alpha_S \leq 0.3$, neglecting quark masses. Specifically, $\zeta$ was found to scale with $\alpha_S^2$ for massless QCD, where conformal symmetry is broken by the running of the coupling $\beta(\alpha_S) \sim \alpha_S^2$:

\begin{equation}
\zeta \sim \frac{T^3}{\alpha_S^2 \log[1/\alpha_S]}\times(\alpha_S^2)^2 \sim \frac{\alpha_S^2 T^3}{\log[1/\alpha_S]} \,\,for\,\, m_0 \ll \alpha_S T.
\end{equation}

The presence of quark masses also leads to breaking conformal invariance provided that $m_0 \leq T$, and this time

\begin{equation}
\zeta \sim \frac{T^3}{\alpha_S^2 \log[1/\alpha_S]}\times (\frac{m_0^2}{T^2})^2 \sim \frac{m_0^4}{T\alpha_S^2 \log[1/\alpha_S]} \,\,for\,\, \alpha_S T \ll m_0 \ll T.
\end{equation}

Here $m_0$ refers to the heaviest zero-temperature quark mass which is smaller than or of order of the temperature T. Only the case of weakly coupled QCD, with vanishing or negligible chemical potential $\mu \sim 0$ was considered. Based on the above results, we can safely conclude that bulk viscosity is negligibly small (at least with respect to shear viscosity) in the perturbative regime of QCD. Specifically:

\begin{equation}\label{param}
\frac{\zeta}{s} \sim \alpha_S^2 \,\,{\rm and}\,\, \frac{\eta}{s} \sim \frac{1}{\alpha_S^2}.
\end{equation}

\begin{figure}
\centering
\includegraphics[width=8cm]{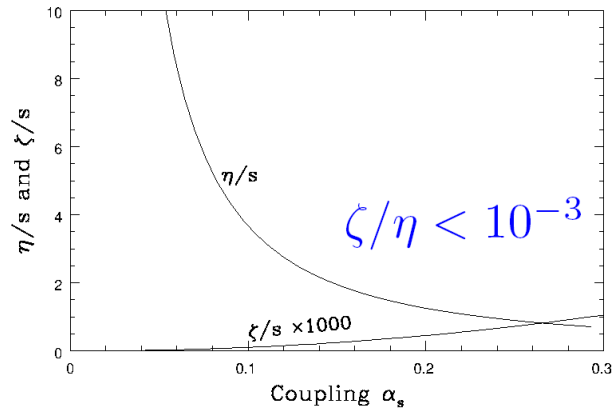}
\caption{Shear versus bulk viscosity: $\eta$/s and $\zeta$/s (s the entropy density) as a function of $\alpha_S$, for $N_f$=3 QCD, neglecting quark masses. Bulk viscosity $\zeta$ has been rescaled by a factor of 1000; from \cite{ADM06}.}
\label{fig:Eta_zeta}
\end{figure}

Note that the parametric dependence of the shear and bulk viscosities on the coupling constant \eq{param} can be easily read off the Kubo formulas \eq{eta_Kubo} and \eq{zeta_Kubo} if we recall the dependence of the QCD action and the trace of the energy-momentum tensor on the coupling constant, see \eq{qcdlagr} and \eq{dilcurr}.

\vskip0.3cm

But what happens at strong coupling, where perturbation theory is no longer applicable? Such is the regime of interest for the Quark-Gluon Plasma. Lattice calculations~\cite{NS04, Mey0704} indicate that shear viscosity gets small, with $\eta$/s not much higher than the conjectured lower bound of 1/4$\pi$ (refer to Sect.~\ref{shearvisc}). Does this mean that the bulk viscosity at strong coupling may become large? 

\vspace{10pt}
Since the bulk viscosity is related by the Kubo formula \eq{zeta_Kubo} to the correlation function 
of the trace of the energy-momentum tensor, this quantity is intimately related to the breaking of scale invariance. In Sect.~\ref{QCDGrav}, we introduced the low-energy theorems for the correlation functions of the trace of the energy-momentum tensor which do not rely on perturbation theory. Let us see what can be said about the bulk viscosity on the basis of this approach; we will follow Ref. ~\cite{KT0705}.

\vskip0.3cm

In this approach, bulk viscosity can be related to the "\textit{interaction measure}", i.e. the expectation value of the trace of the energy-momentum tensor $\left\langle \theta \right\rangle = \epsilon - 3P$, where $\epsilon$ is the energy density and P is the pressure, both of which are measured with high precision on the lattice. Following the definitions and conventions of~\cite{LP80}, we can re-write Eq.~(\ref{zeta_Kubo}) by means of the retarded Green's function as

\begin{equation}
\label{zeta_Green}
\zeta = \frac{1}{9}\lim_{\omega\to 0}\frac{1}{\omega}\int_0^\infty dt \int d^3r\,e^{i\omega t}\,i\,G^R(x) = -\frac{1}{9}\lim_{\omega\to 0}\frac{1}{\omega}\,Im\,G^R(\omega,\stackrel{\rightarrow}{0}),
\end{equation}

where P-invariance was imposed to yield the last expression. Since the spectral density is defined as

\begin{equation}
\label{rho_omega}
\rho(\omega,\stackrel{\rightarrow}{p}) = -\frac{1}{\pi}\,Im\,G^R(\omega,\stackrel{\rightarrow}{p}),
\end{equation}

we can further express the retarded Green's function by using the Kramers-Konig relation to yield~\cite{KT0705}

\begin{equation}
\label{GR_omega}
G^R(\omega,\stackrel{\rightarrow}{p}) = \frac{1}{\pi}\int_{-\infty}^\infty \frac{Im\,G^R(u,\stackrel{\rightarrow}{p})}{u-\omega-i\epsilon} du = \int_{-\infty}^\infty \frac{\rho(u,\stackrel{\rightarrow}{p})}{\omega-u+i\epsilon} du.
\end{equation}

Now, relate by analytic continuation the retarded Green's function for a bosonic excitation to the corresponding Euclidean Green's function

\begin{equation}
G^E(\omega,\stackrel{\rightarrow}{p}) = -G^R(i\omega,\stackrel{\rightarrow}{p}),\, \omega > 0.
\end{equation}

and use~(\ref{GR_omega}) together with the fact that $\rho(\omega,\stackrel{\rightarrow}{p})$ is odd w.r.t. $\omega$ to obtain

\begin{equation}
\label{GE_rho}
G^E(0,\stackrel{\rightarrow}{0}) = 2\int_0^\infty \frac{\rho(u,\stackrel{\rightarrow}{0})}{u} du.
\end{equation}

We can now use the set of low-energy theorems ~\cite{NSVZ81} satisfied by the correlation functions of the trace of the energy-momentum tensor $\theta$ introduced in Sect.~\ref{QCDGrav} to relate the Euclidean Green's function to the thermal expectation value of $\theta$. Following the generalization to the case of finite temperature of~\cite{EKT98}, we get:

\begin{equation}
\label{GE_theta}
G^E(0,\stackrel{\rightarrow}{0}) = \int d^4x \left\langle T\theta(x),\theta(0)\right\rangle = (T\frac{\partial}{\partial T}-4) \left\langle \theta \right\rangle.
\end{equation}

As discussed before in Sect.~\ref{breakscale}, the scale symmetry of the QCD lagrangian is broken by quantum fluctuations, which makes $\theta$ acquire a non-zero expectation value. We now know from Eq.~(\ref{zeta_Kubo}) that a non-zero bulk-viscosity is associated with $\left\langle \theta \right\rangle \neq 0$, fact clearly observed on the lattice for SU(3) gluodynamics~\cite{BEK96} (the same holds in the presence of quarks or at large $N_c$). So let us relate this thermal expectation value to the quantities computed on the lattice via

\begin{equation}
\label{theta_lat}
\left(\epsilon - 3P\right)_{LAT} = \left\langle \theta \right\rangle_T - \left\langle \theta \right\rangle_0 \,\,{\rm with}\,\, \left\langle \theta \right\rangle_0 = -4\left|\epsilon_v\right|,
\end{equation}

where the zero-temperature expectation value of $\theta$, related to the vacuum energy density $\epsilon_v < 0$ has to be subtracted. Using Eq.~(\ref{GE_rho}), (\ref{GE_theta}), (\ref{theta_lat}) we derive an exact sum rule for the spectral density $\rho$ \cite{KT0705}:

\begin{equation}
\label{sum_rule}
2\int_0^\infty \frac{\rho(u,\stackrel{\rightarrow}{0})}{u} du = (T\frac{\partial}{\partial T}-4) \left\langle \theta \right\rangle_T = T^5 \frac{\partial}{\partial T} \frac{\left(\epsilon - 3P\right)_{LAT}}{T^4} + 16\left|\epsilon_v\right|.
\end{equation}

Before extracting the bulk viscosity from~(\ref{sum_rule}), one needs to make an ansatz for the spectral density in the low-frequency regime only -- since the perturbative, divergent contribution has already been subtracted in the definition of the quantities on the r.h.s. of the sum rule~(\ref{sum_rule}). In order to satisfy Eq.~(\ref{zeta_Green}) and~(\ref{rho_omega}), the following ansatz has been assumed in \cite{KT0705}:

\begin{equation}
\label{rho_ansatz}
\frac{\rho(\omega,\stackrel{\rightarrow}{0})}{\omega} = \frac{9\zeta}{\pi} \frac{\omega_0^2}{\omega_0^2+\omega^2}.
\end{equation}

Upon substituting Eq.~(\ref{rho_ansatz}) into~(\ref{sum_rule}) one obtains the much-sought expression for the bulk viscosity:

\begin{equation}
\label{zeta_result}
\zeta = \frac{1}{9\omega_0}\left\{T^5 \frac{\partial}{\partial T} \frac{\left(\epsilon - 3P\right)_{LAT}}{T^4} + 16\left|\epsilon_v\right|\right\}.
\end{equation}

\vskip0.3cm
We have to emphasize that our result depends crucially on the assumed spectral density. 
The shape of the spectral density (and in particular the presence of a $\sim \omega \delta(\omega)$ term) and the behavior of the relaxation time $\sim \omega_0^{-1}$ in the vicinity of the critical point 
both affect the extracted bulk viscosity.  

\vskip3cm

Contrary to what Kubo's formula~(\ref{zeta_Kubo}) may naively imply, $\zeta$ is linear rather than quadratic in the difference $\epsilon-3P$, which seems to be in agreement with the strong-coupling result obtained for the non-conformal SUSY Yang-Mills gauge plasma studied by~\cite{BBS05}. The parameter $\omega_0(T) \propto T$ is the scale at which perturbation theory becomes valid, i.e. when the lattice calculations of the running coupling coincide with the perturbative expression at a given temperature. In the region $1 < \frac{T}{T_c} < 3$ it was found that $\omega_0(T) \approx 1.4 \left(\frac{T}{T_c}\right) GeV$.

\vspace{10pt}
Applying Eq.~(\ref{zeta_result}) to the lattice data from~\cite{BEK96}, with $\left|\epsilon_v\right| = 0.62\,T_c^4$ and $T_c = 0.28~GeV$, the bulk viscosity can be extracted and the ratio $\zeta$/s computed as a function of temperature. It turns out that $\zeta$ is indeed small at high T away from the critical temperature $T_c$ - in agreement with the perturbative results of~\cite{ADM06}, but becomes very large at T close to $T_c$, as confirmed by recent lattice calculations of~\cite{Mey0710, NS07}. A comparison between the results of~\cite{KT0705} and~\cite{Mey0710} is given in Fig.~\ref{fig:Zeta_lat} below.

\begin{figure}
\centering
\includegraphics[width=8cm]{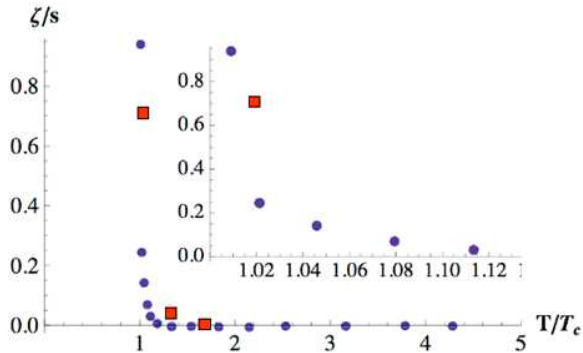}
\caption{$\zeta$/s as a function of T/$T_c$; comparison with lattice results~\cite{Mey0710}.}
\label{fig:Zeta_lat}
\end{figure}

\vspace{10pt}

Corroborated with the lattice results for shear viscosity of~\cite{NS04, Mey0704}, the present result suggests that bulk viscosity may be the dominant correction to ideal hydrodynamics in the vicinity of the deconfinement phase transition in the plasma. Several condensed matter systems, such as $He^3$ near the critical liquid-vapor point, exhibit an analogous behavior, with a large ratio of $\zeta$/$\eta$ affecting sound propagation in these media~\cite{KM98}.

\vspace{10pt}
The analysis described above has been extended  to the case of QCD with 2+1 quark flavors~\cite{KKT0711}, with qualitatively similar results. This latter case is directly relevant for heavy-ion collision experiments, where the two light flavors "up" and "down", along with strangeness, are produced most abundantly. More recent, high statistics lattice data on the equation of state for QCD with almost physical quark masses from the RIKEN-BNL-Columbia-Bielefeld collaboration~\cite{Che07} are used as an input to extract the bulk viscosity.

\vspace{10pt}
The behavior of bulk viscosity near the critical temperature in the presence of light quarks has been discussed in the framework of the effective lagrangians, see \cite{Paech:2006st,Chen:2007kx,FernandezFraile:2008vu,NoronhaHostler:2008ju,Sasaki:2008um}. It is interesting to note that bulk viscosity can exhibit a rapid growth in the vicinity of the chiral critical point \cite{KKT0711}.

\vspace{10pt}
We should mention that the effects of conformal symmetry breaking on bulk viscosity have been studied before in the framework of the gauge-gravity duality, for non-conformal supersymmetric mass-deformed\footnote{Assumes the same mass for all fermions and bosons in the theory.} $\cal{N}$=$2^{\ast}$ Yang-Mills theory in the regime of large 't Hooft coupling $g_{YM}^2 N_c \gg 1$ in~\cite{BBS05}, yielding a linear dependence of $\zeta$ in $\epsilon-3P$ which is similar to~\cite{KT0705}. Further progress has been made by~\cite{GNP0804} and~\cite{GPR0806} in computing bulk viscosity by considering various classes of black hole solutions, which are gravity duals of gauge theories with broken conformal invariance, via the AdS/CFT correspondence introduced in section~\ref{AdSCFT}. The latter results also yield a rise of $\zeta$ in the vicinity of $T=T_c$, though much less sharp than the one predicted in~\cite{KT0705} and~\cite{KKT0711}.

\subsubsection{Bulk viscosity and the mechanism of hadronization.}

Let us now briefly discuss the connection between the growth of bulk viscosity near the critical temperature and an increase in entropy, manifested in abundant particle production in heavy-ion collisions. Namely, the expansion of QCD matter close to the phase transition, produced in such collisions, is accompanied by the production of many soft partons, which screen the color charges of the quarks and gluons present in the medium ~\cite{KKT0711}. Such a scenario may be  called "\textit{soft statistical hadronization}", since the produced partons carry low momenta and the hadronization pattern is not expected to depend on the phase-space distributions of the original partons.

\begin{figure}
\centering
\includegraphics[height=4cm]{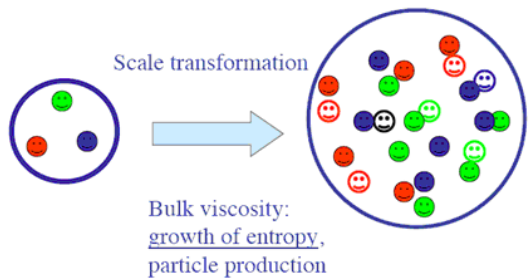}
\caption{The mechanism of soft statistical hadronization as an indicator of entropy growth.}
\label{fig:Entropy}
\end{figure}

\vspace{10pt}
The association of inherent entropy to the hadronization process may be similar to the "\textit{black hole hadronization}" picture that associates an event horizon to the confinement of colored particles~\cite{Kharzeev:2005iz,CKS07}. Since quantum tunnelling turns out to be the only allowed means of crossing the event horizon of quarks, one could in principle relate entropy growth and hadron production to a succession of quantum tunnelling processes that lead to the emission of thermal radiation, which is the QCD counterpart of the Hawking-Unruh radiation emitted by black holes. 

\vspace{10pt}
Within this framework, the results of~\cite{KT0705} and~\cite{Mey0710} shown in Fig.~\ref{fig:Zeta_lat} point towards the $\zeta/s$ vs.~T dependence as a clear indicator of confinement, as seen by the off-equilibrium thermodynamics. In heavy-ion collisions, this may be manifested through both a decrease of the average transverse momentum of the resulting particles and an increase in the total particle multiplicity.  Let us also mention an interesting scenario 
\cite{Torrieri:2008ip} where the growth of bulk viscosity induces an instability in the hydrodynamical flow of the plasma.

\section {Limitations of the present approaches and Outlook.}

In these lectures, we have introduced the two dualities relating QCD and gravity, and discussed  applications to the computation of the shear and bulk viscosities of strongly-coupled QCD matter. However useful and mathematically sound these dual approaches may seem, there are some serious drawbacks associated with the use of these methods  which we outline below.

\vspace{10pt}
In case of the AdS/CFT correspondence presented in section~\ref{AdSCFT}, it relates perturbative string theory calculations to nonperturbative (strong coupling) calculations in the 4 dimensional $\cal N$ = 4 Super Yang-Mills theory, which are otherwise very difficult to obtain. The ultimate interest is, however, to perform strong coupling calculations in the real world --- in QCD, the theory of the strong interaction. $\cal N$ = 4 Super Yang-Mills is quite far from QCD, in particular by being supersymmetric and conformally invariant. The fact that AdS/CFT is defined for a gauge group SU($N_c$) as a perturbation around $N_c = \infty$ further complicates matters.

\vspace{10pt}
In case of the second method, one major but hopefully solvable problem is to include the quarks 
in the effective action. It may also be useful to formulate the effective action explicitly as a gravity 
action in five dimensions and to study the duality of solutions of the effective dilaton-gluon action on one side and of the Einstein-Hilbert gravity action on the other. The string--like confining solution 
of the effective dilaton-gluon action has been obtained recently in \cite{Kharzeev:2008br} where it was found that the formation of the string is accompanied by the emergence of a massless dilaton mode. 

\vspace{10pt}
The two approaches outlined in these lectures are at first glance completely different: it is enough to mention that one is essentially guided by conformal invariance whereas the other is driven by the pattern of conformal invariance {\it breaking}. Nevertheless,   
it is quite likely that once the breaking of scale invariance is introduced within the AdS/CFT correspondence on one hand (see \cite{Karch:2006pv,Gursoy:2007cb,GNP0804} for work in this direction), and the effective dilaton action is formulated 
in dual language as a five-dimensional gravity on the other, 
the two approaches may appear closer than it seems at present.
In any case, even a remote prospect of finding some day a gravity dual of QCD 
certainly justifies giving it a try. 

\vskip0.3cm

D.K. is grateful to H. Satz and B. Sinha for the invitation to deliver these lectures and the hospitality in Jaipur. The work of D.K. was supported by the U.S. Department of Energy under Contract No. DE-AC02-98CH10886.

\printindex

\end{document}